\documentclass[12pt]{iopart}
\usepackage{iopams}
\usepackage{enumitem}
\usepackage{graphicx}
\usepackage{cite}
\usepackage{braket}

\newcommand{\aver}[1]{\langle #1 \rangle}
\newcommand{\thetaa}{\theta_{\textsc{\scriptsize a}}}
\newcommand{\thetab}{\theta_{\textsc{\scriptsize b}}}
\newcommand{\thetaat}{\theta_{\textsc{\scriptsize a}}}
\newcommand{\thetabt}{\theta_{\textsc{\scriptsize b}}}
\newcommand{\smtextsc}[1]{\textsc{\scriptsize #1}}
\newcommand{\vardel}{\Delta}
\graphicspath{{./figures/}}
\usepackage{xcolor}

\newcommand{\epsarg}[1]{\varepsilon_{\small\textsc{\scriptsize #1}}}

\begin{document}

\title{Tomographic entanglement indicators in frequency combs and Talbot carpets}

\author{B. Sharmila$^{1,2}$, S. Lakshmibala$^{1}$, and V. Balakrishnan$^{1}$}

\address{$^{1}$ Department of Physics, Indian Institute of Technology Madras, Chennai 600036, India. \\
$^{2}$ Present address: Department of Physics, University of Warwick, Coventry CV4 7AL, UK.}
\vspace{10pt}
\begin{indented}
\item[]\today
\end{indented}

\begin{abstract}
Recent theoretical investigations on tomographic entanglement indicators have showcased the advantages of the tomographic approach in the context of continuous-variable (CV), spin and hybrid quantum systems. Direct estimation of entanglement using experimental data from the IBM quantum computing platform and NMR experiments have also been carried out in earlier work. A similar investigation in the context of CV systems is necessary to fully assess the utility of our tomographic approach. In this paper, we highlight the advantages of our approach in the context of experiments reported in the literature on two CV systems, namely, entangled Talbot carpets and biphoton frequency combs. We use the tomographic entanglement indicator to assess the extent of entanglement between a pair of Talbot carpets and demonstrate that this provides a simpler and more direct procedure compared to the one suggested in the experiment. We also establish that the tomograms corresponding to the two biphoton frequency combs carry clear entanglement signatures that distinguish between the two states. 
\end{abstract}
\vspace{2pc}

\maketitle
%
%
\section{\label{sec:intro}Introduction}

The measurement of any  observable in  a quantum mechanical system yields a histogram of the state of the system  in the basis of that observable. Measurements of a judiciously chosen \textit{quorum} of appropriate observables of a system that are informationally complete, yield a set of histograms called a tomogram.
Quantum state reconstruction seeks to obtain the density matrix from the tomogram. 
However, reconstruction of the state of a radiation field from the corresponding optical tomogram could be both tedious and complex~\cite{recon2019}. In many contexts, the reconstruction program is also limited by the presence of inherent aberrations in the source~\cite{recon2020, recon2018}. It is worth noting that even in the simple case of a two-level atom interacting with a radiation field, the state of the field subsystem was experimentally reconstructed from the corresponding tomogram at various instants of temporal evolution only as recently as 2017~\cite{ReconJCM2017}. With an increase in the number of field modes interacting with an atomic system, the inevitable entanglement that arises during dynamical evolution makes state reconstruction a more formidable task.  
It would 
therefore  be efficient to \textit{read off} information about a state, wherever possible, directly from the experimentally accessible histograms. In particular, estimating the extent of entanglement through simple manipulations of the relevant tomograms \textit{alone} becomes an interesting and important exercise. 

Entanglement is an essential resource in quantum information processing, and assessing the extent of entanglement between two subsystems of a bipartite system is necessary for this purpose. 
One of the standard measures of entanglement between the two subsystems A and B of a bipartite system is the subsystem von Neumann entropy $\xi_{\smtextsc{svne}}=-\mathrm{Tr}\,(\rho_{i} \,\log_{2} \,\rho_{i})$ where $\rho_{i}$ ($i=\mathrm{A,\,B}$) is the reduced density matrix of the subsystem concerned. 

The computation of $\xi_{\smtextsc{svne}}$, however, requires a knowledge of the full density matrix, in contrast to the tomographic approach. The advantage of the latter has been showcased in certain continuous-variable (CV) systems as,  for instance, in identifying qualitative signatures of entanglement between the radiation fields in the output ports of a quantum beam-splitter~\cite{sudhrohithbs}. Further, detailed analysis of entanglement has been carried out theoretically using several indicators obtained directly from tomograms in different contexts, 
e.g., close to avoided energy-level crossings~\cite{sharmila4}. Their efficacy has been assessed by comparison with standard entanglement measures~\cite{sharmila,sharmila2}. Further, these indicators have been examined using the IBM quantum computing platform by translating certain multipartite hybrid quantum (HQ) systems into suitable equivalent circuits~\cite{sharmila3}. It is to be noted that in these examples, measurements of multiple observables corresponding to the rotated quadrature operators 
\begin{equation}
X_{\theta}= (a e^{-i \theta} + a^{\dagger} e^{i \theta})/\sqrt{2}
\end{equation}
(where $(a,a^{\dagger})$ are the photon annihilation and creation operators and $\theta\in[0,\pi)$) are, in principle, required. In stark contrast, this is not the case in certain recent experiments reported in the literature as, for instance, in an experiment~\cite{QTalbCarp} involving an entangled pair of Talbot carpets generated from photons produced from spontaneous parametric down conversion (SPDC). Here an entanglement indicator $I_{\smtextsc{d}}$ based on Bell-type inequalities is used. It is interesting to note that the Talbot carpets have also been  used in other quantum contexts 
such as  information processing~\cite{milmanPRATalbot}. Another interesting example~\cite{milman} pertains to biphoton frequency combs. Qubits are encoded in this CV system using the time-frequency continuous degrees of freedom of photon pairs generated through the SPDC process. Such photon pairs have ben 
 studied  extensively for investigating entanglement properties and potential logic gate operations~\cite{milmanPRA2020,milman2020}.

In such experiments, the tomographic approach mentioned earlier provides an efficient, readily usable procedure for capturing signatures and understanding the qualitative features of bipartite entanglement, as will be shown in what follows. For instance, even in the experiment pertaining to a pair of entangled Talbot carpets, the proposed experimental setup can be simplified by adopting the tomographic approach. A phase shifter, which is necessary in the proposed experimental arrangement for $I_{\smtextsc{d}}$, is not needed for estimating the extent of entanglement using a tomographic entanglement indicator. Recently, it has been demonstrated~\cite{TalbotExpt}  
that it is possible to implement single-qudit logic gate operations using the Talbot effect. 
The next step would be to extend this investigation to the case of a pair of entangled Talbot carpets generated using a suitable combination of two such experimental setups. It is here that the tomographic approach comes into its own, as it has the advantage of simplifying considerably the necessary experimental setup. Further, in both experimental situations under discussion, one involving entangled Talbot carpets and the other involving biphoton frequency combs, we demonstrate how convenient and useful the tomographic indicators are.
The versatility of the tomographic approach is brought out by the fact that optical tomograms are useful in the experiment on entangled Talbot carpets, while chronocyclic tomograms display the features of entanglement in the experiment on biphoton frequency combs. 

The plan of this paper is as follows. In \sref{sec:revIndics}, we review the salient features of the relevant tomograms, and describe the entanglement indicators that can be obtained directly from these tomograms. In \sref{sec:TalbCarp}, we apply our procedure to assess the extent of entanglement between a pair of Talbot carpets. Whereas it has been suggested in~\cite{QTalbCarp} that an experimental setup which uses Bell-type inequalities could determine the extent of entanglement, we demonstrate that tomograms provide a simpler and more direct procedure to assess entanglement. Further, in \sref{sec:biphoton}, we establish that the respective chronocyclic tomograms corresponding to the two biphoton states provide manifest entanglement signatures that distinguish between the states. We conclude the paper  with some brief remarks.

\section{\label{sec:revIndics} Tomograms}

\subsection{Quadrature histograms}

We begin by describing the salient features of the optical tomograms. In a two-mode CV system, the infinite set of rotated quadrature operators~\cite{VogelRisken, LvovskyRaymer, ibort} given by
\begin{equation}
\label{eqn:quadop}
\mathbb{X}_{\thetaat} = (a^\dagger e^{i \thetaat} + a e^{-i \thetaat})/\sqrt{2}, \qquad \mathrm{and} \qquad \mathbb{X}_{\thetabt} = (b^\dagger e^{i \thetabt} + b e^{-i \thetabt})/\sqrt{2}
\end{equation}
where  $\thetaa, \, \thetab \in [0,\pi)$,  constitute the quorum of observables that carries complete information about the state.
 Here $(a, a^{\dagger})$ and $(b, b^{\dagger})$ are the annihilation and creation operators corresponding  respectively to subsystems A and B of the bipartite system, so that 
 $[a, a^{\dagger}] = 1$ and $[b, b^{\dagger}] = 1$. The bipartite tomogram is 
\begin{equation}
w(X_{\thetaat}, \thetaa; X_{\thetabt}, \thetab)= \aver{X_{\thetaat}, \thetaa; X_{\thetabt}, \thetab|\rho_{\smtextsc{ab}}|X_{\thetaat}, \thetaa; X_{\thetabt}, \thetab},
\label{eqn:2modeTomoDefn}
\end{equation}
where $\rho_{\smtextsc{ab}}$ is the bipartite density matrix and $\mathbb{X}_{\theta_{i}} \ket{X_{\theta_{i}},\theta_{i}} = X_{\theta_{i}} \ket{X_{\theta_{i}},\theta_{i}}$ ($i=\mathrm{A,B}$). Here, 
$\ket{X_{\thetaat},\thetaat;X_{\thetabt},\thetabt}$  stands for 
$\ket{X_{\thetaat},\thetaat} \otimes \ket{X_{\thetabt},\thetabt}$. 
The normalisation condition is given by 
\begin{equation}
\int_{-\infty}^{\infty}\! \rmd X_{\thetaat} \int_{-\infty}^{\infty} \!\rmd X_{\thetabt} w(X_{\thetaat},\thetaa;X_{\thetabt},\thetab) = 1
\label{eqn:tomoNorm}
\end{equation}
for each $\thetaa$ and $\thetab$. We note that $\thetaa=\thetab=0$ corresponds to the position quadrature. As we are only interested  
in the corresponding slice of the tomogram, we adopt the following simplified notation. A slice of the tomogram (equivalently, the histogram in the position basis)  for a bipartite state 
$\ket{\psi_{\smtextsc{ab}}}$ is given by
\begin{equation}
w(x_{\smtextsc{a}}; x_{\smtextsc{b}}) = |\aver{x_{\smtextsc{a}};x_{\smtextsc{b}}| \psi_{\smtextsc{ab}}}|^{2}
\label{eqn:xslice}
\end{equation}
Here, $\lbrace \ket{x_{\smtextsc{x}}} \rbrace$ ($\textsc{x=a,b}$) are the position eigenstates in the two  subsystems,  and 
$\ket{x_{\smtextsc{a}};x_{\smtextsc{b}}}$ denotes the corresponding factored product state. The marginal distributions obtained from the joint distribution above yield the corresponding slices $w_{\smtextsc{a}}(x_{\smtextsc{a}})$ and 
$w_{\smtextsc{b}}(x_{\smtextsc{b}})$ of the subsystem tomograms (reduced tomograms). 

\subsection{Chronocyclic tomogram}

The analogy between an ultrashort light pulse and a quantum mechanical wave function 
leads~\cite{paye} to a \textit{chronocyclic} representation for the  study of  ultrashort pulses,  
where the time $t$ and frequency $\omega$ are the conjugate observables.
The state of a single photon of 
frequency $\omega$ is denoted in a spectral representation of infinitely narrow-band pulses
 by $\ket{\omega}$.  The superposed state of a photon that has a frequency $\omega$ with a probability amplitude $\mathcal{S}(\omega)$ is given by $\int \rmd \omega \, \mathcal{S}(\omega) \ket{\omega}$. In an equivalent temporal representation of infinitely short-duration  pulses $\{\ket{t}\}$, this $1$-photon state is $\int \rmd t \, \widetilde{\mathcal{S}}(t) \ket{t}$, where $\widetilde{\mathcal{S}}(t)$ is the Fourier transform of $\mathcal{S}(\omega)$.
A family of rotated observables $(\omega \cos \theta + t \sin \theta)$ can then be defined~\cite{beck}, where $\omega$ and $t$ 
have been  scaled by a natural time scale of the 
system to make them dimensionless. Measurements of these rotated observables form the basis of chronocyclic 
tomography, in which  the set of histograms corresponding to these observables gives the tomogram of the state of a $1$-photon system.
In this chronocyclic representation, a $1$-photon state  can  also be described  in the time-frequency `phase space' by a corresponding Wigner function~\cite{paye}. 
Extension to multipartite states corresponding to two or more photons is straightforward. For instance, two photons of frequencies $\omega$ and $\omega^{\prime}$ 
are given by the 
biphoton CV  bipartite state $\ket{\omega} \otimes \ket{\omega^{\prime}}$. 

In the experiment with biphoton frequency combs, what is of relevance to us is the time-time slice of the bipartite chronocylic tomogram corresponding to state $\ket{\psi}$, given by,
\begin{equation}
w(t;t^{\prime})=|\aver{t;t^{\prime} |\psi}|^{2},
\label{eqn:tslice}
\end{equation}
where  $\ket{t;t^{\prime}}$ 
stands for $\ket{t}\otimes\ket{t^{\prime}}$ in the temporal representation. This is the chronocyclic analogue of~\eref{eqn:xslice}. The marginal distributions are obtained from~\eref{eqn:tslice} as described before.

\subsection{Tomographic entanglement indicators}

The extent of  correlation between the subsystems can be deduced from the tomographic entropies. For instance, for the slice defined in~\eref{eqn:xslice}, the bipartite tomographic entropy is given by 
\begin{equation}
S_{\smtextsc{ab}} = - \int_{-\infty}^{\infty} \, \rmd x_{\smtextsc{a}} \int_{-\infty}^{\infty} \, \rmd x_{\smtextsc{b}} w(x_{\smtextsc{a}};x_{\smtextsc{b}})\, \log_{2} \, w(x_{\smtextsc{a}};x_{\smtextsc{b}}).
\label{eqn:2modeEntropy}
\end{equation}
The subsystem tomographic entropy is
\begin{equation}
S_{i} = - \int_{-\infty}^{\infty} \: \rmd x_{i} \: w_{i}(x_{i}) \log_{2} \,w_{i}(x_{i}) \;\; (i= \textsc{A,B}).
\label{eqn:1modeEntropy}
\end{equation}
Signatures of the extent of entanglement can be gleaned from the mutual information~\cite{sharmila},
\begin{equation}
\epsarg{tei}=S_{\smtextsc{a}} + S_{\smtextsc{b}} - S_{\smtextsc{ab}}.
\label{eqn:epsTEI}
\end{equation}
In the chronocyclic case, analogous definitions of entropies and mutual information hold for any tomographic slice 
given by ~\eref{eqn:tslice}.

\section{\label{sec:TalbCarp}Entangled Talbot carpets}

We first review, in brief, 
 the salient features of the proposed experimental setup pertaining to entangled Talbot carpets~\cite{QTalbCarp}. We then compute $\epsarg{tei}$ corresponding to the tomographic slice~\eref{eqn:xslice}. We emphasize that this procedure circumvents state reconstruction. Finally, we compare the performance of $\epsarg{tei}$ 
 vis-\`a-vis other measures. 

\begin{figure}
\includegraphics[width=0.55\textwidth]{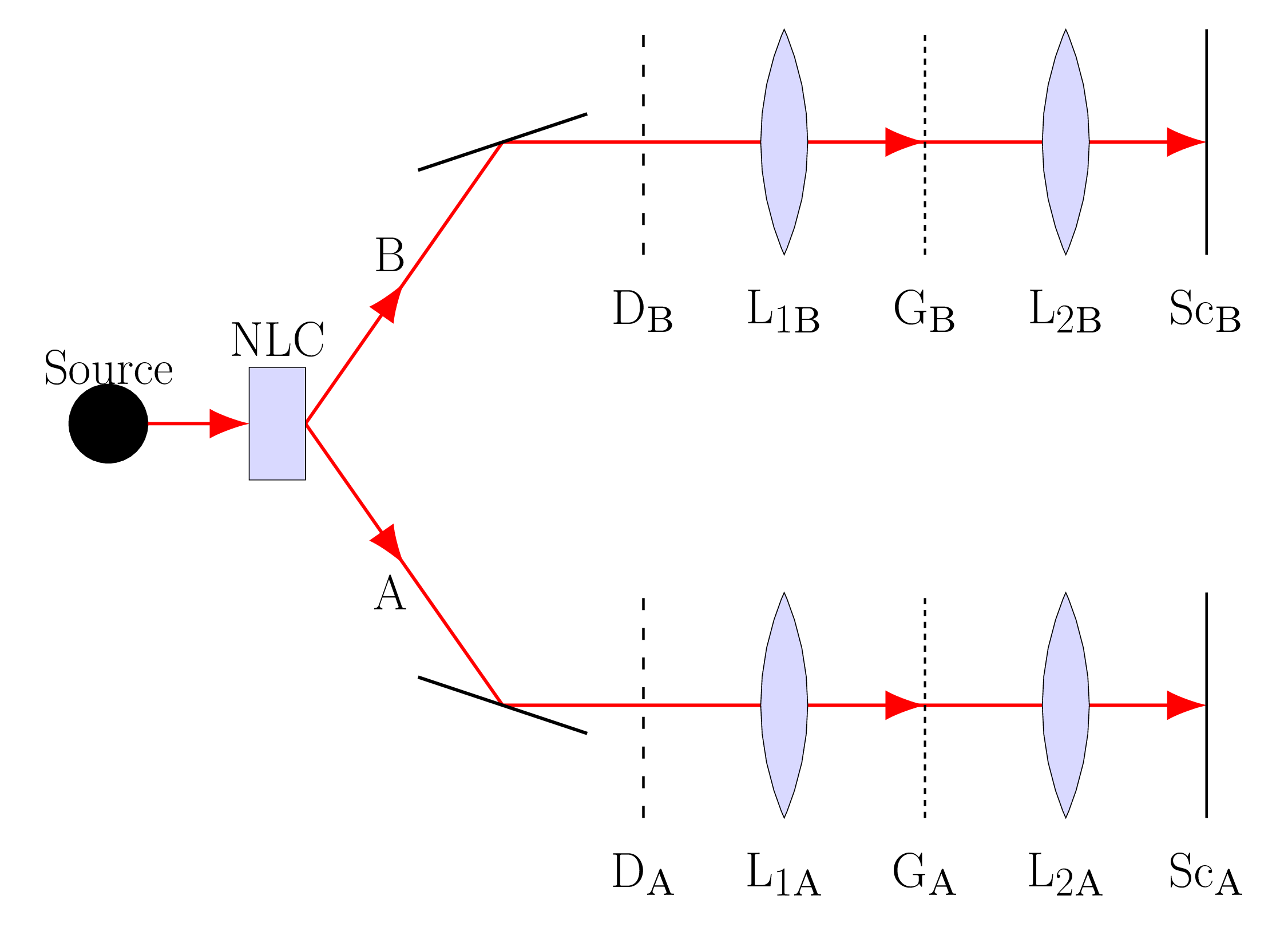}
\caption{Experimental setup: A pump photon from the laser source is incident on a nonlinear crystal (NLC) producing two SPDC photons, each of which passes through a $D$-slit aperture ($\mathrm{D}_{i}$), a lens ($\mathrm{L}_{1 i}$), a grating ($\mathrm{G}_i$), and another lens ($\mathrm{L}_{2 i}$). $\mathrm{Sc}_{i}$ is the detection screen ($i=\textsc{a,b}$).}
\label{fig:setup}
\end{figure}

In the proposed experiment (\Fref{fig:setup}), light from a laser source passes through a nonlinear crystal (NLC). Entangled SPDC photon pairs are produced, with a spatial correlation 
given by 
\begin{equation}
\nonumber R=\frac{\kappa_{+}^{2} - \kappa_{-}^{2}}{\kappa_{+}^{2} + \kappa_{-}^{2}} = -\frac{\Delta_{+}^{2}}{\Delta_{-}^{2}}.
\end{equation}
Here $\kappa_{+}$ is the width of the pump field frequency profile, $\kappa_{-}$ is the standard deviation in the phase matching of the two output photons, and 
\begin{equation}
\nonumber \frac{1}{\Delta_{\pm}^{2}} = \frac{1}{\kappa_{+}^{2}} \pm \frac{1}{\kappa_{-}^{2}}. 
\end{equation}
The light is then guided along two different paths A and B using appropriate mirrors. Along path $i$ ($i=\textsc{a,b}$), a $D$-slit aperture $\mathrm{D}_{i}$, a lens $\mathrm{L}_{1 i}$, a grating $\mathrm{G}_{i}$, a lens $\mathrm{L}_{2 i}$ and a screen $\mathrm{Sc}_{i}$ are placed as shown on the figure. Talbot carpets are seen on each screen, and the extent of entanglement between the two are to be assessed. The slit width in each aperture is $\delta$, and the inter-slit spacing is $s$. Each grating has slit width $\sigma$ and period $\ell$. The screens are in the $x-y$ plane.

The entangled Talbot state is of the form
\begin{equation}
\ket{\Psi}=\sum_{d_{1},d_{2}=0}^{D-1} C_{d_{1},d_{2}} \ket{d_{1}}_{\smtextsc{a}} \otimes \ket{d_{2}}_{\smtextsc{b}},
\label{eqn:CarpSt_defn}
\end{equation}
where
\begin{equation}
\nonumber C_{d_{1},d_{2}}=\mathcal{N} \exp\Big\{-\frac{s^{2}}{4 \Delta_{+}^{2}} \left(d_{1}^{2} - 2 R d_{1} d_{2} + d_{2}^{2} \right)
\Big\},
\end{equation}
where $\mathcal{N}$  is the normalization constant such that 
$\sum\limits_{d_{1},d_{2}} |C_{d_{1},d_{2}}|^{2} =1$. The basis states are now given by 
\begin{equation}
\aver{x_{i}|d}_{i}=T_{d}(x_{i})= \mathcal{A}_{d} \sum_{m=-\infty}^{\infty} \exp \Big\{-\frac{(2 \pi m \sigma)^{2}}{2 \ell^{2}}- \frac{(x_{i} - d s - m \ell)^{2}}{4 \delta^{2}}\Big\}, \: 
\label{eqn:basis}
\end{equation}
where  $i=\textsc{a,b}$ and 
$\mathcal{A}_{d}$ is the normalisation constant. It is straightforward to measure the intensity distribution at different points $(x_{\smtextsc{a}},\, x_{\smtextsc{b}})$ along the $x$-axis on the two screens. The expression for this distribution can be obtained from $\Psi(x_{\smtextsc{a}},\, x_{\smtextsc{b}})$ by using \eref{eqn:basis}.

The extent of entanglement has been assessed from $I_{\smtextsc{d}}$ for different values of the spatial correlation $R$ and the number of slits $D$~\cite{QTalbCarp}. For this purpose, two pairs of local measurements $A_{1}$, $A_{2}$ (respectively, $B_{1}$, $B_{2}$) are performed on both A and B. Each of these four measurements has $D$ outcomes with projective operators $\lbrace \ket{f_{\alpha_{1}}}_{\smtextsc{a \, a}}\hspace*{-0.25 em}\bra{f_{\alpha_{1}}} \rbrace$, $\lbrace \ket{f_{\alpha_{2}}}_{\smtextsc{a \, a}}\hspace*{-0.25 em}\bra{f_{\alpha_{2}}} \rbrace$, $\lbrace \ket{g_{\beta_{1}}}_{\smtextsc{b \, b}}\hspace*{-0.25 em}\bra{g_{\beta_{1}}} \rbrace$, and $\lbrace \ket{g_{\beta_{2}}}_{\smtextsc{b \, b}}\hspace*{-0.25 em}\bra{g_{\beta_{2}}} \rbrace$ corresponding to $A_{1}$, $A_{2}$, $B_{1}$, and $B_{2}$ respectively. Here
\begin{eqnarray}
\ket{f_{\alpha_{j}}}_{\smtextsc{a}}&=\frac{1}{\sqrt{D}}\sum_{d=0}^{D-1} e^{2 \pi i d (f+\alpha_{j})/D}\ket{d}_{\smtextsc{a}},\\
\ket{g_{\beta_{j}}}_{\smtextsc{b}}&=\frac{1}{\sqrt{D}}\sum_{d=0}^{D-1} e^{2 \pi i d (-g+\beta_{j})/D}\ket{d}_{\smtextsc{b}} \; (j=1,2)
\label{eqn:phaseShifter}
\end{eqnarray}
with $\alpha_{1}=0$, $\alpha_{2}=0.5$, $\beta_{1}=0.25$, $\beta_{2}=-0.25$ and $f,g=0,1,\dots,D-1$. (In the tomographic approach that will be outlined later, these phase shifts $\alpha_{j}$, $\beta_{j}$ ($j=1,2$) need not be implemented, and the entanglement indicator can be deduced solely from the original intensity patterns on the screens). 
It can be seen~\cite{cglmp} that for $I_{\smtextsc{d}}\leq 2$, the states are unentangled. Here,
\begin{equation}
I_{\smtextsc{d}}= \sum_{k=0}^{[D/2]-1} \left( 1 - \frac{2 k}{D-1} \right) J_{k},
\end{equation}
where
\begin{eqnarray*}
J_{k}=&\:P(A_{1}=B_{1}+k)\, -  \,P(A_{1}=B_{1}-k-1) \\
&+P(B_{2}=A_{1}+k)\, - \,P(B_{2}=A_{1}-k-1) \\
&+P(B_{1}=A_{2}+k+1)\, - \,P(B_{1}=A_{2}-k)\\
&+P(A_{2}=B_{2}+k)\, - \,P(A_{2}=B_{2}-k-1)
\end{eqnarray*}
and
\begin{equation}
\nonumber P(A_{i}=B_{j}+k)=\sum_{p=0}^{D-1} P(A_{i}=(p+k) \, \mathrm{mod} \, D, B_{j}=p).
\end{equation}
The joint probability distribution of the outcomes being $A_{i}=p$ and $B_{j}=q$ is denoted by $P(A_{i}=p,B_{j}=q)$ .


\begin{figure}
\includegraphics[width=0.4\textwidth]{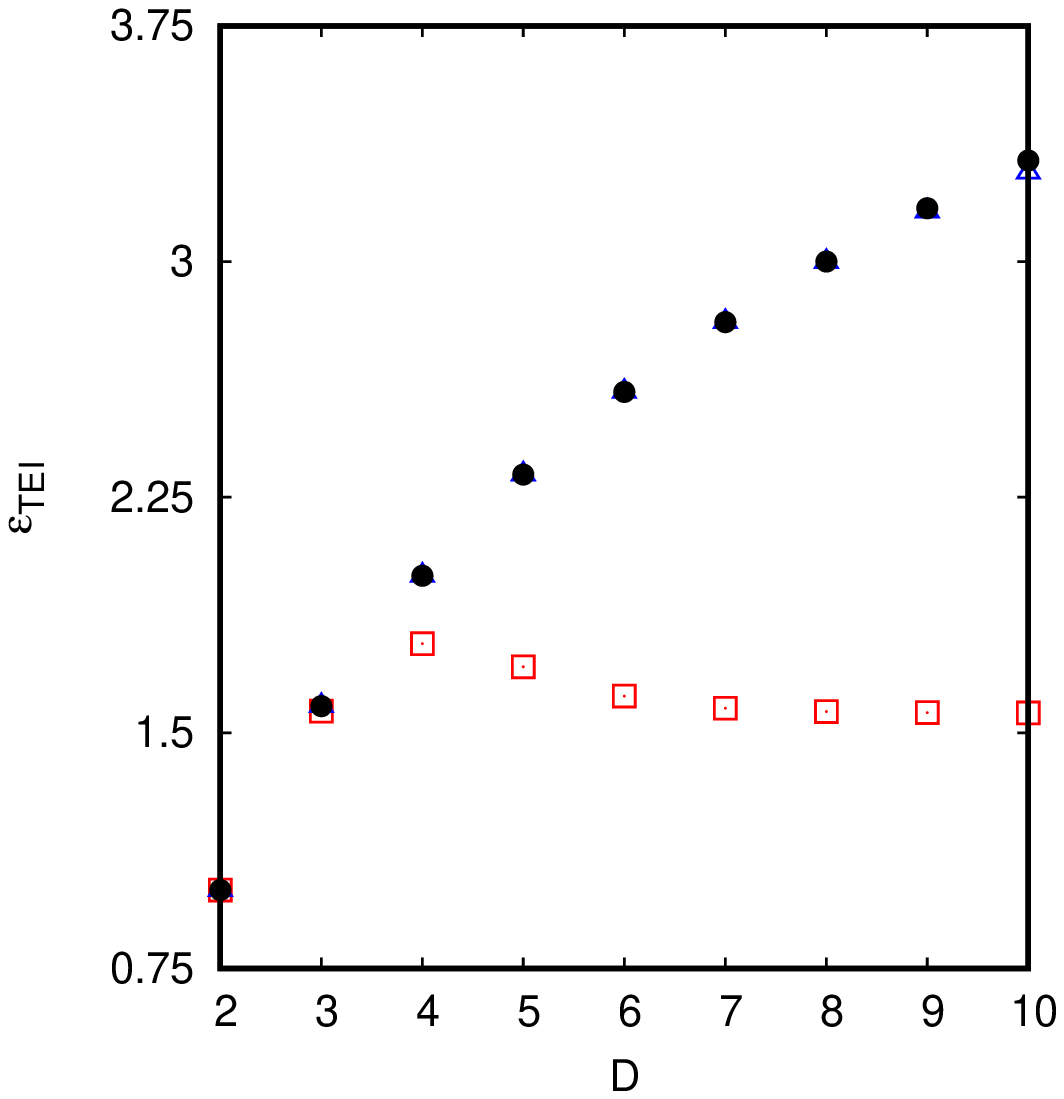}
\includegraphics[width=0.4\textwidth]{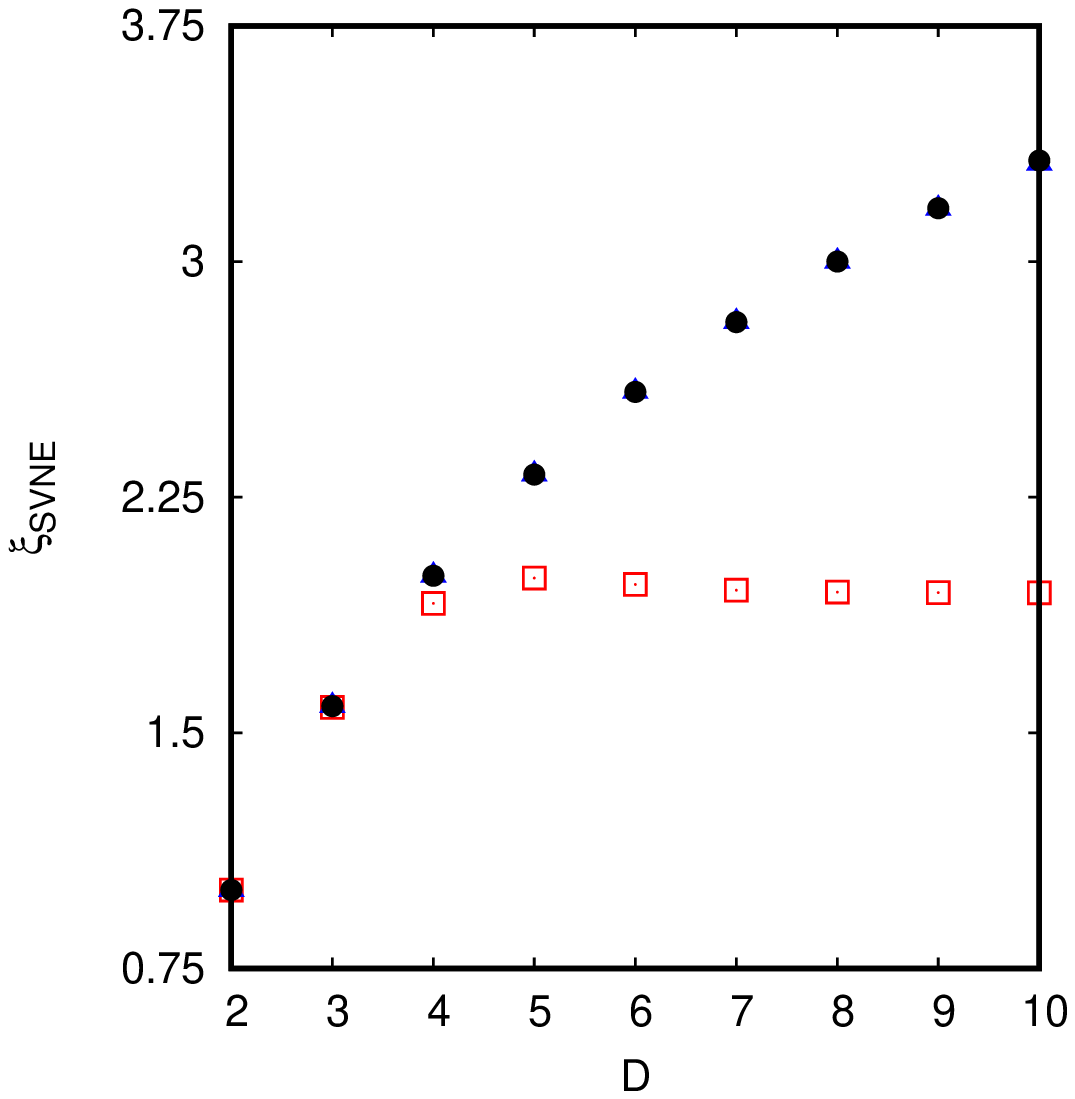}\\
\includegraphics[width=0.4\textwidth]{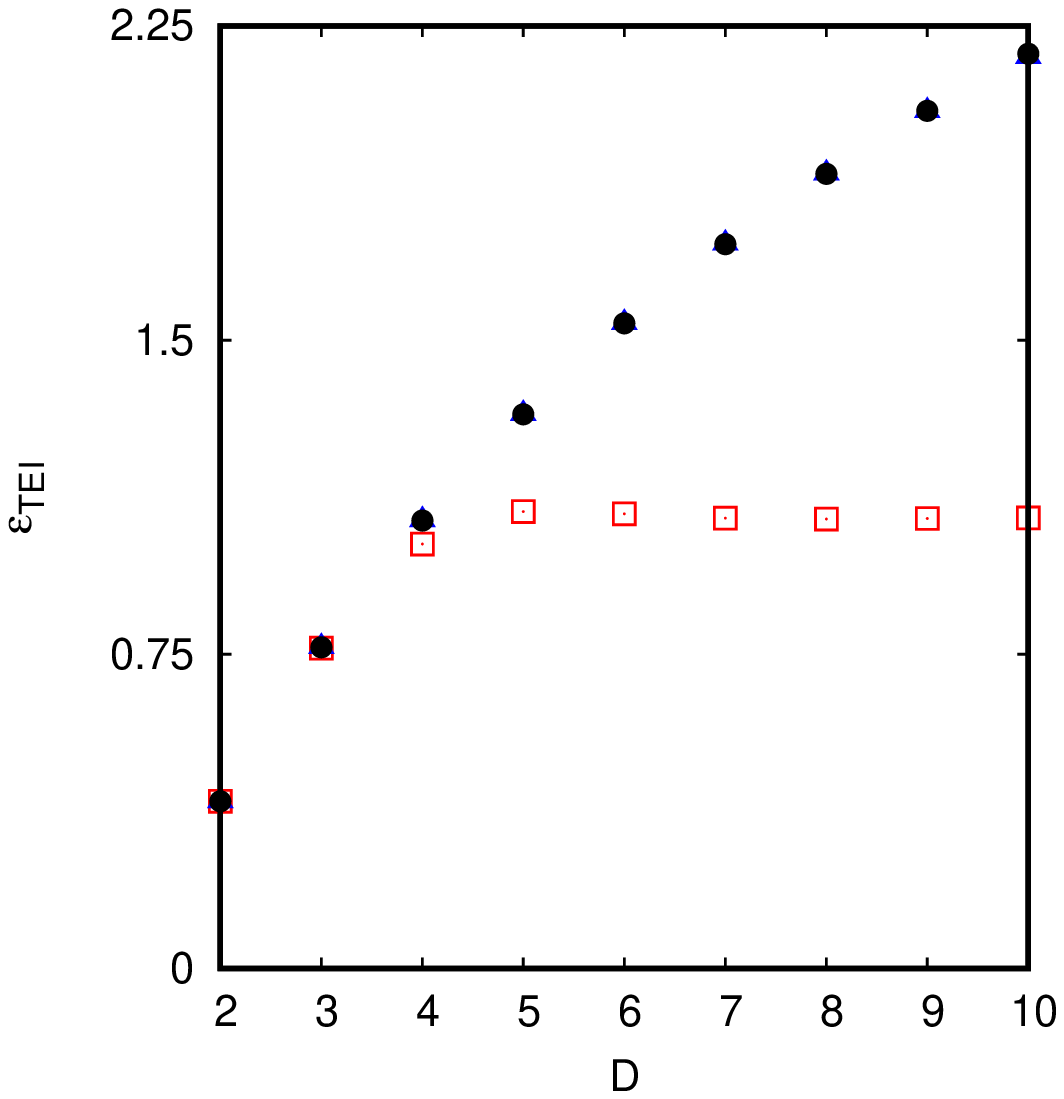}
\includegraphics[width=0.4\textwidth]{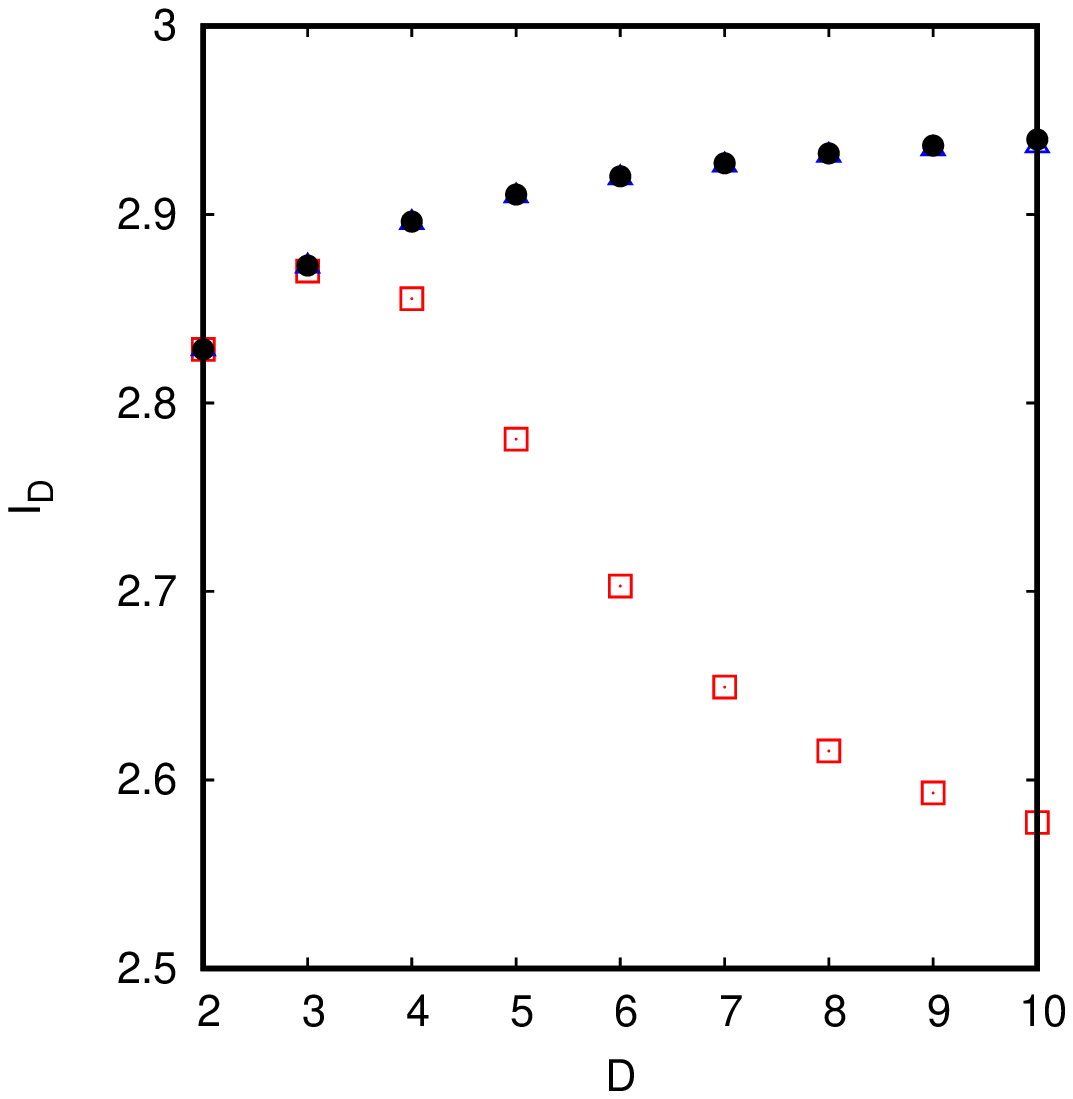}
\caption{$\epsarg{tei}$ computed in the $x_{\smtextsc{a}}-x_{\smtextsc{b}}$ basis (top left), $\xi_{\smtextsc{svne}}$ (top right), $\epsarg{tei}$ computed in the $A_{1}-B_{1}$ basis (bottom left),  and $I_{\smtextsc{d}}$ (bottom right) vs $D$ for $0.9998$ (red), $0.99998$ (blue), and $1$ (black). Bottom right figure courtesy 
Ref.  \cite{QTalbCarp}.}
\label{fig:compare_tei_Talb_2}
\end{figure}

\begin{figure}
\includegraphics[width=0.32\textwidth]{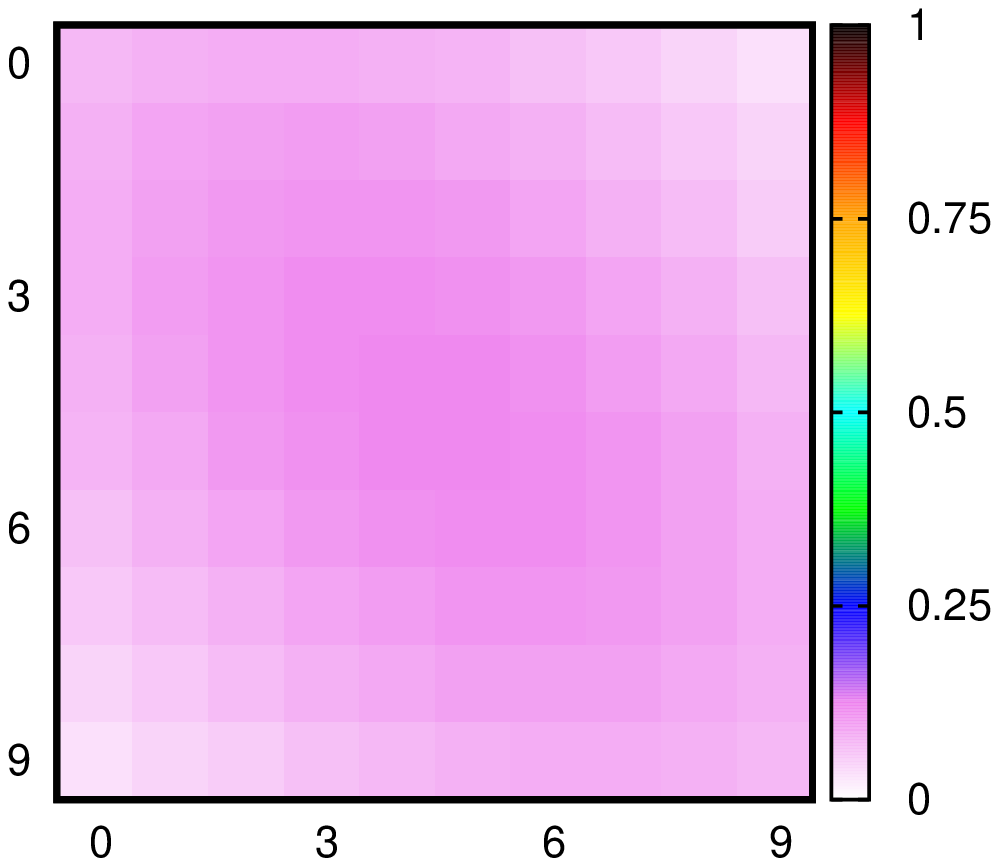}
\includegraphics[width=0.32\textwidth]{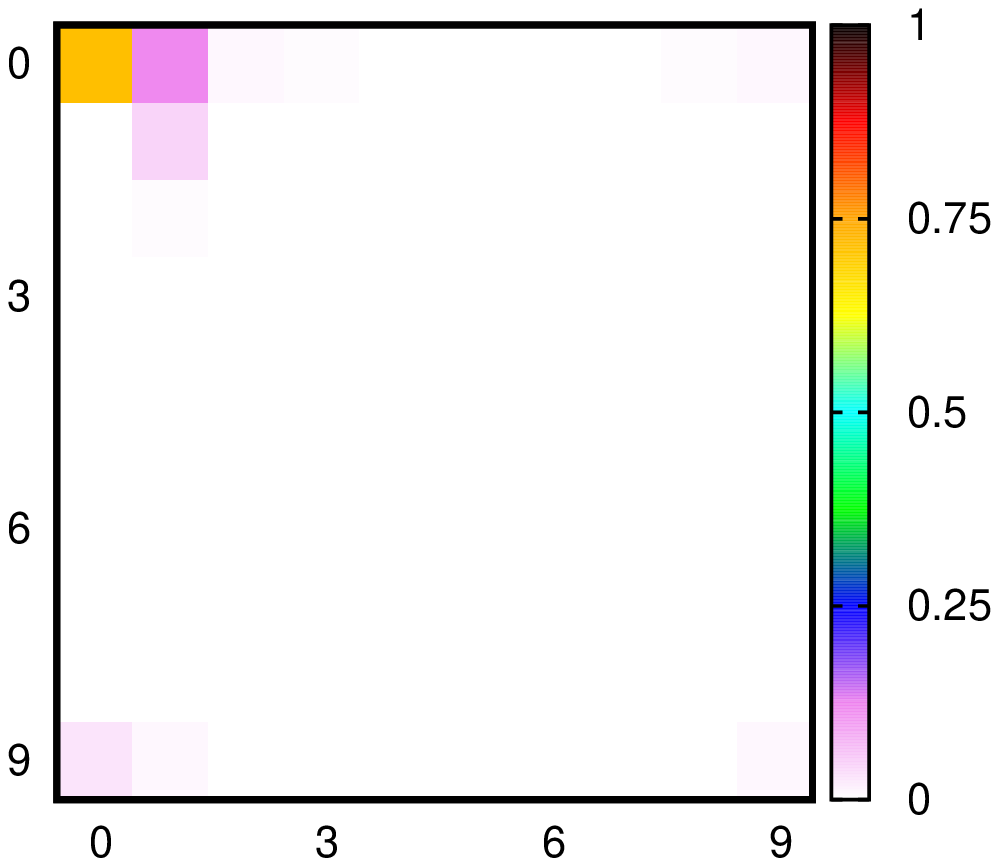}\\
\includegraphics[width=0.32\textwidth]{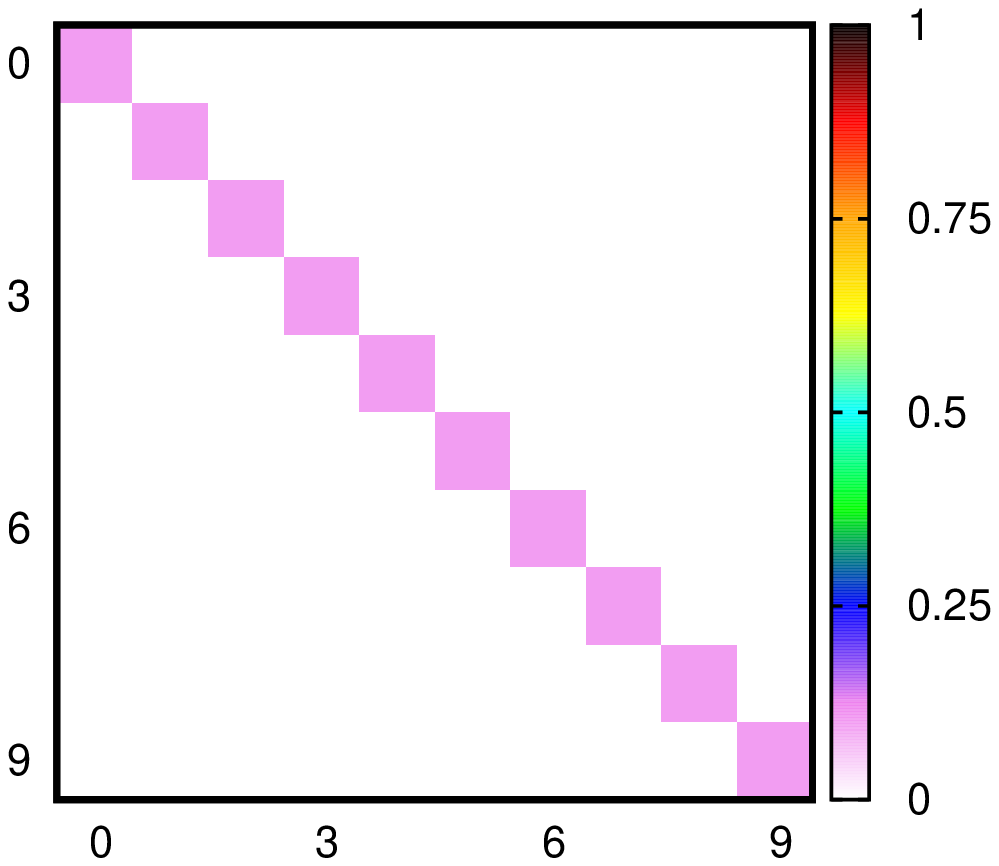}
\includegraphics[width=0.32\textwidth]{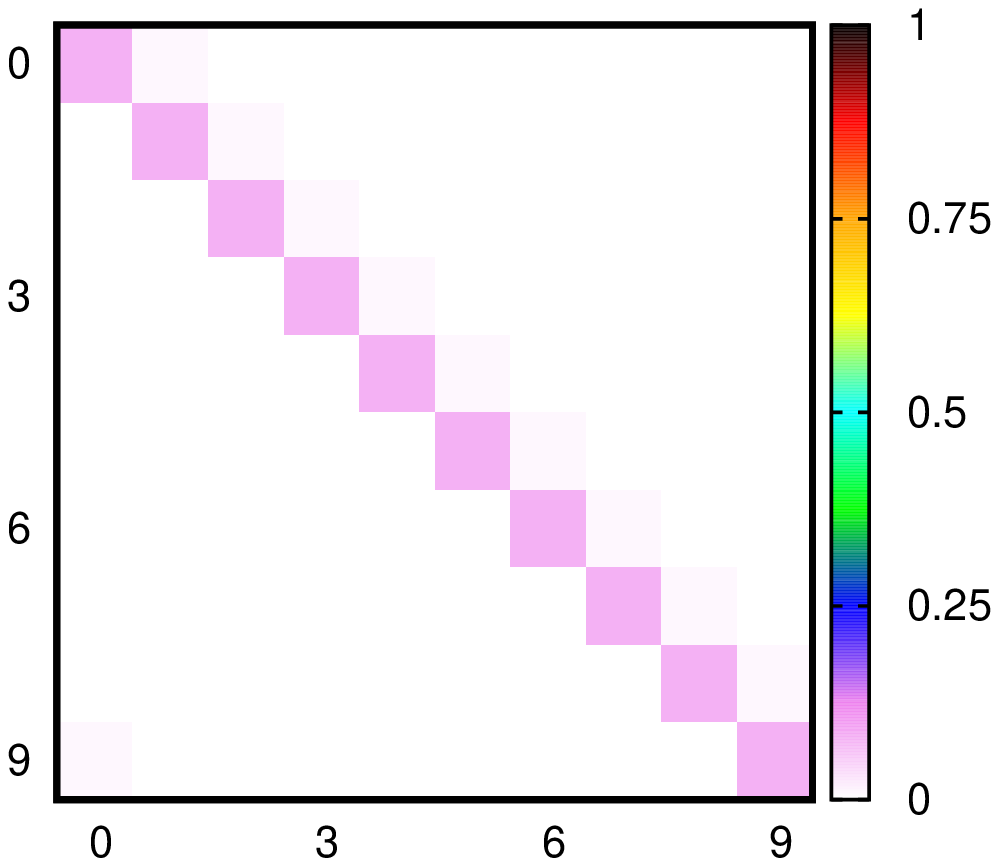}
\caption{Density matrix corresponding to subsystem A for $R=0.998$ (top left) and $R=1$ (bottom left). Respective joint probability distributions $P(A_{1}=p,B_{1}=q)$ (right panel) vs $p$ and $q$.}
\label{fig:explainID}
\end{figure}

In the tomographic approach (as in the case of $I_{\smtextsc{d}}$), we set $\sigma=0.05 \ell$, $\delta=0.025 s$, $\ell=1$, $s=1/D$, and $\kappa_{+}=9 \ell$ to facilitate comparison. As outlined in \Sref{sec:revIndics}, we compute $\epsarg{tei}$ (see \eref{eqn:epsTEI}) from the optical tomogram \eref{eqn:xslice}.
In \Fref{fig:compare_tei_Talb_2}, $\epsarg{tei}$, $\xi_{\smtextsc{svne}}$, and $I_{\smtextsc{d}}$ is plotted versus $D$ for various $R$ values. It is clear that $\epsarg{tei}$ agrees well with $\xi_{\smtextsc{svne}}$. Further, $\epsarg{tei}$ captures the gross features of $I_{\smtextsc{d}}$. 
From \Fref{fig:compare_tei_Talb_2}, we see that when $\epsarg{tei}$ is computed from tomograms corresponding to other basis, the results do not change, and that the extent of entanglement significantly increases even with very small changes  in $R$. (For $D=10$, $\epsarg{tei}\sim 0.3$ for $R=0.998$, and $\sim 3.32$ for $R=1$ correct to two decimal places). This follows from the fact that the state for $R=1$ is far more entangled than that for $R=0.998$ as can be seen from the off-diagonal elements of the subsystem density matrices (\Fref{fig:explainID}).

\section{\label{sec:biphoton}Biphoton frequency combs}

We first recapitulate the salient features of the experiment~\cite{milman} on two pairs of entangled biphoton states, denoted by $\ket{\Psi_{\alpha}}$ 
and $\ket{\Psi_{\beta}}$, which are frequency 
combs comprising finite-width  peaks.
%
The entangled photons were generated using SPDC. In the setup, the resonant frequency of the cavity is denoted by  
$\overline{\omega}$, and the pump photons have frequency $\omega_{p}$. If  
$\omega_{\smtextsc{s}}$ and 
$\omega_{\smtextsc{i}}$  
denote the signal 
and idler frequencies, respectively, 
and $\Omega$ is their difference, it can be seen (Appendix A) that $\ket{\Psi_{\alpha}}$ 
and $\ket{\Psi_{\beta}}$ can be expressed as given below.
\begin{eqnarray}
\nonumber 
\ket{\Psi_{\alpha}}
= \mathcal{N}_{\alpha}^{-1/2}\int \rmd \omega_{\smtextsc{s}} \int \rmd \omega_{\smtextsc{i}} &f_{+}(\omega_{\smtextsc{s}}+\omega_{\smtextsc{i}}) f_{-}(\Omega)
\times \\
 & f_{\mathrm{cav}}(\omega_{\smtextsc{s}}) f_{\mathrm{cav}}(\omega_{\smtextsc{i}}) \ket{\omega_{\smtextsc{s}}} \otimes \ket{\omega_{\smtextsc{i}}}.
\label{eqn:Psi_plpl}
\end{eqnarray}
Here, 
\begin{equation}
f_{-}(\Omega) = e^{ -(\Omega - \Omega_{0})^{2}/
4 (\vardel \Omega)^{2} },
\label{eqn:fmin}
\end{equation}
where $\Omega_{0}$ 
and $\vardel \Omega$ are the 
mean and standard deviation of $\Omega$. 
$f_{\mathrm{cav}}$ is the Gaussian comb
\begin{equation}
f_{\mathrm{cav}}(\omega)
=\sum_{n} e^{-(\omega -n \overline{\omega})^{2}/
2 (\vardel \omega)^{2}}
\label{eqn:fcav}
\end{equation}
where $\vardel \omega$ is the 
standard deviation of each Gaussian, $\mathcal{N}_{\alpha}$ is the normalisation constant,
and $f_{+}(\omega_{\smtextsc{s}}+\omega_{\smtextsc{i}})=\delta(\omega_{p}-\omega_{\smtextsc{s}}-\omega_{\smtextsc{i}})$.
 We note that~\eref{eqn:Psi_plpl} features the product $f_{\mathrm{cav}}(\omega_{\smtextsc{s}})f_{\mathrm{cav}}(\omega_{\smtextsc{i}})$, where $f_{\mathrm{cav}}(\omega)$  
 is a superposition of Gaussians corresponding to 
 odd and even values of $n$ such that the two are in phase with each other.

The second biphoton state is given by 
\begin{eqnarray}
\nonumber \ket{\Psi_{\beta}}=
\mathcal{N}_{\beta}^{-1/2}\int \rmd \omega_{\smtextsc{s}} \int \rmd \omega_{\smtextsc{i}} &f_{+}(\omega_{\smtextsc{s}}+\omega_{\smtextsc{i}}) f_{-}(\Omega)     
\times \\
 & g_{\mathrm{cav}}(\omega_{\smtextsc{s}}) f_{\mathrm{cav}}(\omega_{\smtextsc{i}}) \ket{\omega_{\smtextsc{s}}} \otimes \ket{\omega_{\smtextsc{i}}}, 
 \label{eqn:Psi_mipl}
\end{eqnarray}
where 
\begin{equation}
g_{\mathrm{cav}}(\omega)
=\sum_{n}  (-1)^{n} e^{-(\omega -n \overline{\omega})^{2}/
2 (\vardel \omega)^{2}}.
\label{eqn:gcav}
\end{equation}
Here, $\mathcal{N}_{\beta}$ is the normalisation constant. 
In contrast to $\ket{\Psi_{\alpha}}$,~\eref{eqn:Psi_mipl} features the product 
$g_{\mathrm{cav}}(\omega_{\smtextsc{s}}) 
f_{\mathrm{cav}}(\omega_{\smtextsc{i}})$, 
where $g_{\mathrm{cav}}(\omega)$ 
is a  superposition of Gaussians corresponding to 
odd and even $n$  such that the two are out of phase with each other. In what follows, we will use the tomographic approach to distinguish between the two biphoton states. Since photon \textit{coincidence} counts were used to experimentally distinguish between the two states, it is reasonable to expect that the time-time slices of the tomograms corresponding to the two biphoton states will capture the difference.
The time-time slices are denoted by
\begin{eqnarray}
w^{\smtextsc{x}}(t_{\smtextsc{s}};t_{\smtextsc{i}})=\aver{ t_{\smtextsc{s}};t_{\smtextsc{i}} \ket{\Psi_{\smtextsc{x}}}\bra{\Psi_{\smtextsc{x}}} t_{\smtextsc{s}};t_{\smtextsc{i}} },
\label{eqn:tt_slice}
\end{eqnarray}
where $(\textsc{x}=\alpha,\beta)$, and $\ket{t_{\smtextsc{s}};t_{\smtextsc{i}}}$ 
stands for $\ket{t_{\smtextsc{s}}}\otimes\ket{t_{\smtextsc{i}}}$ in the time-time basis. 

We work with  the parameter values used 
 in Ref. ~\cite{milman}, namely,  
$\omega_{p}/(2 \pi)=391.8856$ THz, 
$\overline{\omega}/(2 \pi)=19.2$ GHz, 
$\vardel\omega/(2 \pi)=1.92$ GHz, 
$\Omega_{0}/(2 \pi)=10.9$ THz, and 
$\vardel\Omega/(2 \pi)=6$ THz. 
The time-time slices of the tomograms of 
$\ket{\Psi_{\alpha}}$ and $\ket{\Psi_{\beta}}$ have 
been obtained by 
substituting~\eref{eqn:Psi_plpl}  and 
\eref{eqn:Psi_mipl} in turn 
in~\eref{eqn:tt_slice} and simplifying the 
resulting expressions  
(see Appendix B). 

\begin{figure}
\centering
\includegraphics[width=0.32\textwidth]{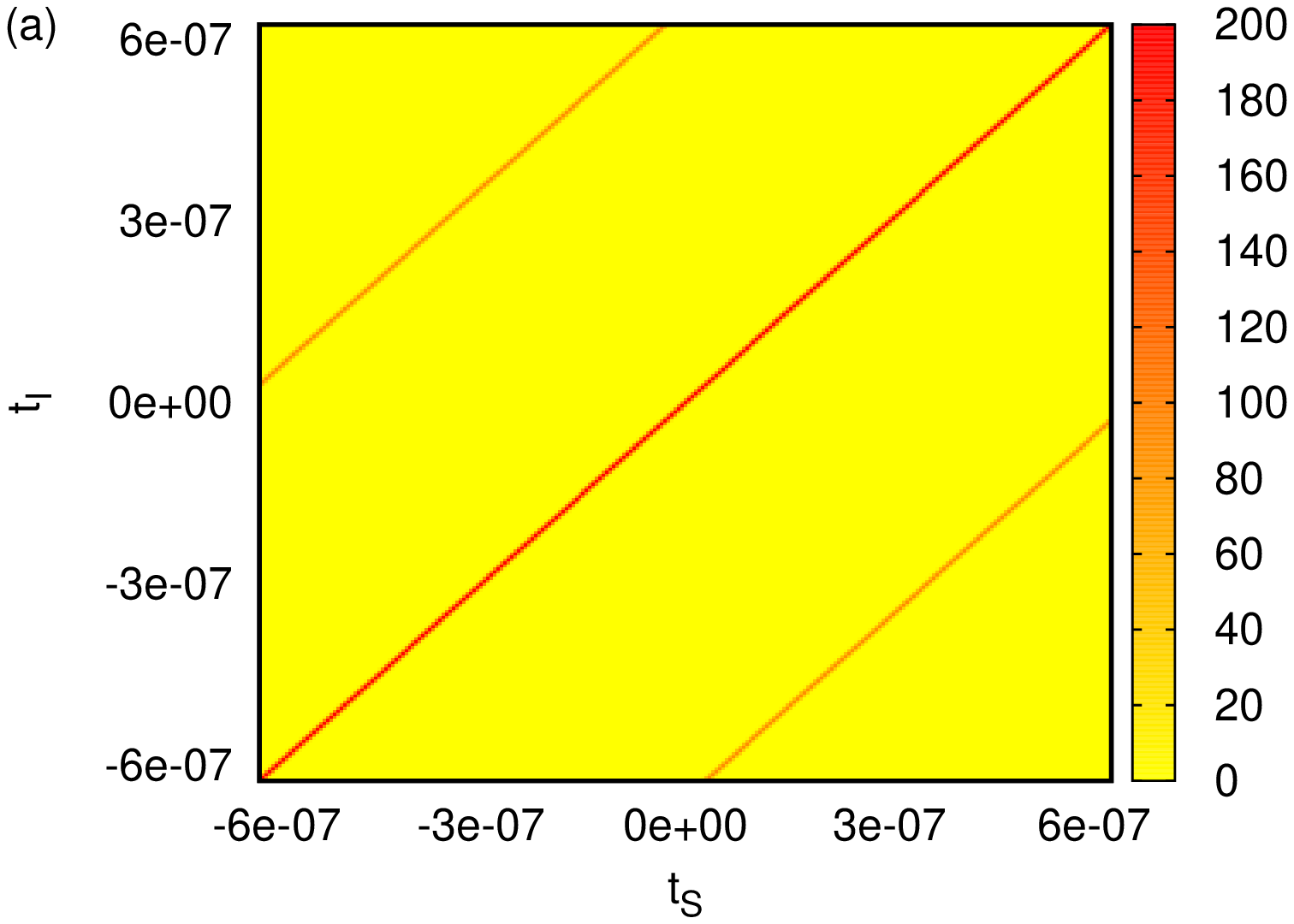}
\includegraphics[width=0.32\textwidth]{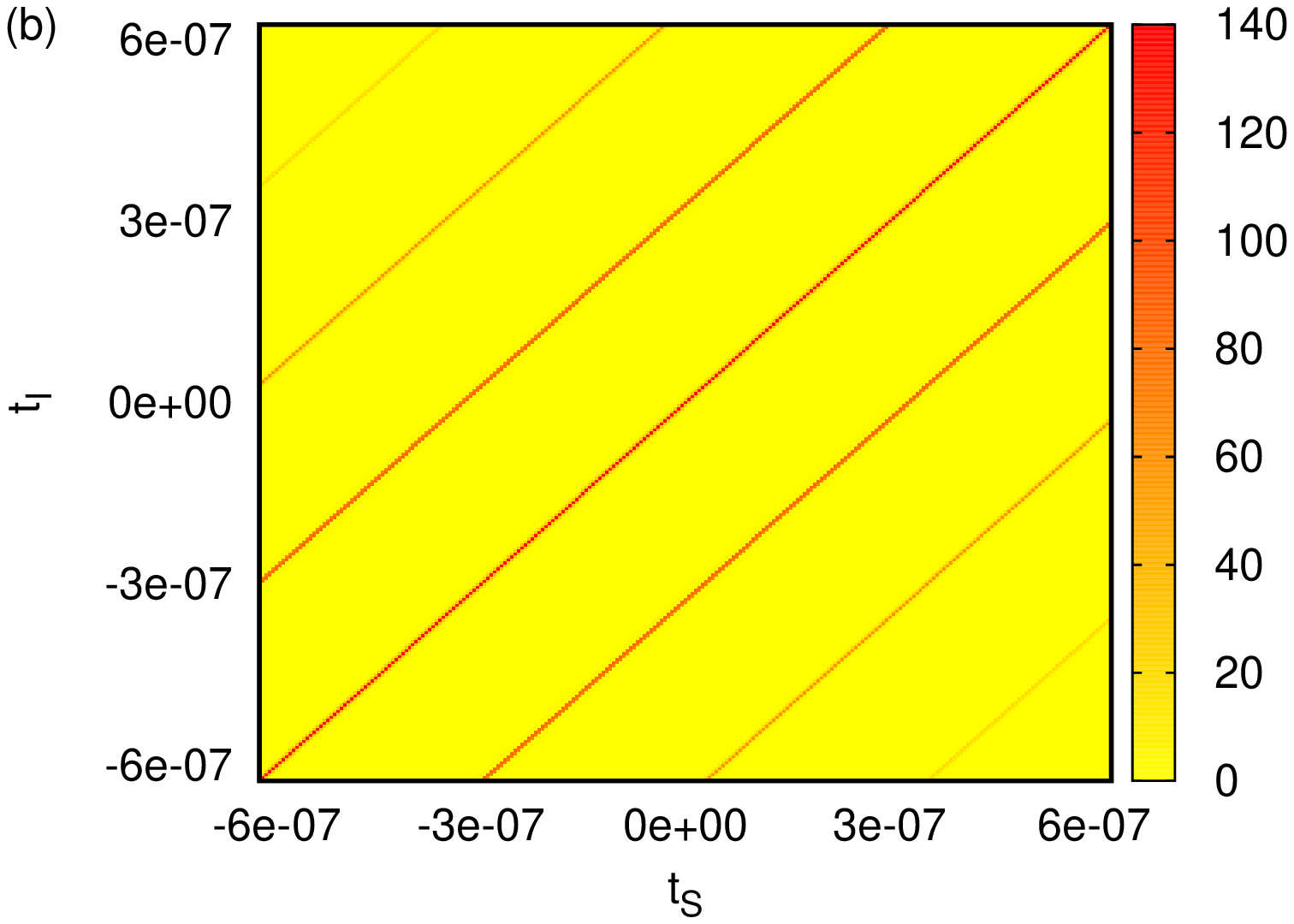}
\includegraphics[width=0.32\textwidth]{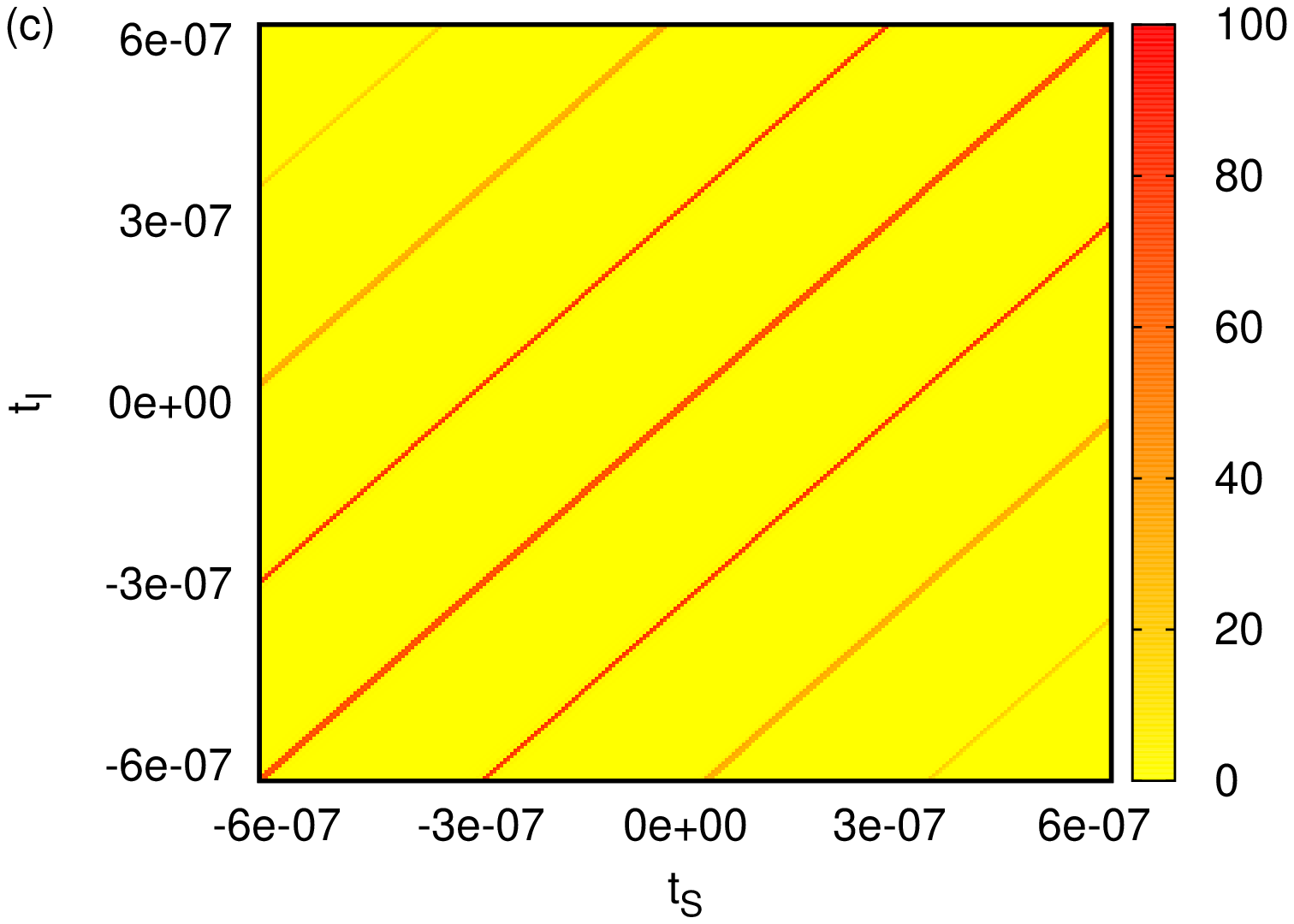}
\caption{Tomographic time-time slice (a) $w^{\alpha}(t_{\smtextsc{s}};t_{\smtextsc{i}})$  and (b) $w^{\beta}(t_{\smtextsc{s}};t_{\smtextsc{i}})$ vs. $t_{\smtextsc{s}}$ and $t_{\smtextsc{i}}$ in seconds. (c) Difference $|w^{\alpha}(t_{\smtextsc{s}};t_{\smtextsc{i}})-w^{\beta}(t_{\smtextsc{s}};t_{\smtextsc{i}})|$ vs. $t_{\smtextsc{s}}$ and $t_{\smtextsc{i}}$ in seconds.}
\label{fig:chrono_tomo}
\end{figure}

As expected, $w^{\alpha}(t_{\smtextsc{s}};t_{\smtextsc{i}})$ and $w^{\beta}(t_{\smtextsc{s}};t_{\smtextsc{i}})$  are distinctly different 
from each other, as seen in  figures~\ref{fig:chrono_tomo} (a)-(c). 
This  difference arises because $\ket{\Psi_{\alpha}}$ and $\ket{\Psi_{\beta}}$ correspond to combs that are clearly displaced with respect to each other, when expressed in the time-time basis.

Next, we calculate  the reduced tomograms 
$w^{\smtextsc{x}}_{i}(t_{i})$ 
corresponding to subsystem $i$ (where  
$i=\textsc{s,i}$ and $\textsc{x} =\alpha,\beta$) 
by integrating out the other subsystem. 
(For instance, $w^{\smtextsc{x}}_{\smtextsc{s}}(t_{\smtextsc{s}})=\int \rmd t_{\smtextsc{i}} w^{\smtextsc{x}}(t_{\smtextsc{s}};t_{\smtextsc{i}})$.)  
Using these full-system and subsystem chronocyclic tomograms in~\eref{eqn:2modeEntropy}--\eref{eqn:epsTEI}, we obtain the entanglement indicator $\epsarg{tei}$ corresponding to any chosen slice of the chronocyclic tomogram.
(For ease of notation, we have dropped the explicit dependence of 
$\epsarg{tei}$ on the choice of both the tomogram slice and the specific state.) 
In the case of the time-time slice of the tomograms 
we get, finally, the values $\epsarg{tei} = 6.50$ 
for  the state $\ket{\Psi_{\alpha}}$, and 
$\epsarg{tei} = 5.44$ for the state  
$\ket{\Psi_{\beta}}$. Thus,  $\epsarg{tei}$ clearly distinguishes between these two biphoton states. 
 We emphasize that the methods used by us could, in principle, provide an alternative approach to the procedure adopted in the experiment.

\section{Concluding remarks}

The tomographic entanglement indicator $\epsarg{tei}$ proves to be a very useful tool which is also easily computed from the histogram of a relevant measured observable. In the case of the entangled Talbot states, $\epsarg{tei}$ computed from the histogram in the position basis, closely mimics the standard entanglement measure $\xi_{\textsc{svne}}$. We emphasize that a single slice suffices to estimate the extent of entanglement, and we do not require the rotated quadratures in this case. In fact, we have shown that $\epsarg{tei}$ is better than $I_{\textsc{d}}$, the Bell-like-inequality-based indicator. Further, we have unambiguously distinguished between a pair of biphoton states using the entanglement indicator $\epsarg{tei}$. This paper demonstrates alternative procedures using the tomographic approach that are useful and efficient in a variety of experimentally relevant CV systems.

\section*{Acknowledgments}
We acknowledge useful discussions with P. Milman, Laboratoire Mat$\acute{\mathrm{e}}$riaux et Ph$\acute{\mathrm{e}}$nom$\grave{\mathrm{e}}$nes Quantiques, Universit$\acute{\mathrm{e}}$ de Paris.
This work was supported in part by a seed grant from IIT Madras to the Centre for Quantum Information, Communication and Computing, under the IoE-CoE scheme. 

\appendix
\section*{\label{appen:2states}Appendix A: The biphoton frequency comb states}
\setcounter{section}{1}

In Section \ref{sec:biphoton}, the expressions for the two biphoton states $\ket{\Psi_{\alpha}}$ and $\ket{\Psi_{\beta}}$ which are distinguished from each other using tomograms are given in~\eref{eqn:Psi_plpl} and \eref{eqn:Psi_mipl} respectively. For convenience, we give the expressions below.
\begin{eqnarray*}
\ket{\Psi_{\alpha}}
= \mathcal{N}_{\alpha}^{-1/2}\int \rmd \omega_{\smtextsc{s}} \int \rmd \omega_{\smtextsc{i}} f_{+}(\omega_{\smtextsc{s}}+\omega_{\smtextsc{i}}) f_{-}(\Omega)
  f_{\mathrm{cav}}(\omega_{\smtextsc{s}}) f_{\mathrm{cav}}(\omega_{\smtextsc{i}}) \ket{\omega_{\smtextsc{s}}} \otimes \ket{\omega_{\smtextsc{i}}},
\end{eqnarray*}
where $\mathcal{N}_{\alpha}$ is the normalisation constant. Here, $f_{-}(\Omega)$ and $f_{\mathrm{cav}}(\omega)$
are defined in~\eref{eqn:fmin} and \eref{eqn:fcav} respectively. 
The other biphoton state
\begin{eqnarray*}
\ket{\Psi_{\beta}}=
\mathcal{N}_{\beta}^{-1/2}\int \rmd \omega_{\smtextsc{s}} \int \rmd \omega_{\smtextsc{i}} f_{+}(\omega_{\smtextsc{s}}+\omega_{\smtextsc{i}}) f_{-}(\Omega)     
 g_{\mathrm{cav}}(\omega_{\smtextsc{s}}) f_{\mathrm{cav}}(\omega_{\smtextsc{i}}) \ket{\omega_{\smtextsc{s}}} \otimes \ket{\omega_{\smtextsc{i}}}, 
\end{eqnarray*}
where $g_{\mathrm{cav}}(\omega)$ is defined in~\eref{eqn:gcav}, 
and $\mathcal{N}_{\beta}$ is the normalisation constant.

In what follows, we outline the procedure to show that these two states are indeed the two states which were shown to be distinguishable using photon coincidence counts, in the experiment reported in 
Ref. \cite{milman}. 

In the experimental setup, the sum of the frequencies of the signal and the idler photons matches the pump frequency, i.e., $(\omega_{\smtextsc{s}}+\omega_{\smtextsc{i}}=\omega_{\smtextsc{p}})$. Hence, as stated in the supplementary material of \cite{milman}, $f_{+}(\omega_{\smtextsc{s}}+\omega_{\smtextsc{i}})$ can be replaced by $\delta(\omega_{\smtextsc{s}}+\omega_{\smtextsc{i}}-\omega_{p})$. Integrating over the variable $\Omega_{+}\:(=\omega_{\smtextsc{s}}+\omega_{\smtextsc{i}})$, appropriately changing the integration variables, noting that $\Omega=\omega_{\smtextsc{s}}-\omega_{\smtextsc{i}}$, and dropping the normalisation constant, we get
\begin{equation}
\ket{\Psi_{\alpha}}
= \int \rmd \Omega f_{-}(\Omega) f_{\mathrm{cav}}(\omega_{\smtextsc{s}}) f_{\mathrm{cav}}(\omega_{\smtextsc{i}}) \ket{\omega_{\smtextsc{s}}} \otimes \ket{\omega_{\smtextsc{i}}}.
\label{eqn:B19milman}
\end{equation}
This can be identified as one of the states considered in the experiment, namely, the expression~(B19) in \cite{milman}, on changing the notation from $\Omega$, $\omega_{\smtextsc{s}}$, $\omega_{\smtextsc{i}}$ in~\eref{eqn:B19milman} to $\omega_{-}$, $\omega_{s}$, $\omega_{i}$ respectively.

We now proceed to establish that the other biphoton state $\ket{\Psi_{\beta}}$ considered by us, is the same as the state $\ket{\psi^{\prime}}\:(=C^{\prime} Z_{t_{\smtextsc{s}}} \ket{\widetilde{+}}_{\omega_{\smtextsc{s}}} \otimes \ket{\widetilde{+}}_{\omega_{\smtextsc{i}}})$ defined in \cite{milman}. Here, $C^{\prime}\ket{t_{\smtextsc{s}};t_{\smtextsc{i}}}=\ket{t_{\smtextsc{s}}+t_{\smtextsc{i}};t_{\smtextsc{s}}-t_{\smtextsc{i}}}$ where, for instance, $\ket{t_{\smtextsc{s}}}\otimes\ket{t_{\smtextsc{i}}}$ is denoted by $\ket{t_{\smtextsc{s}};t_{\smtextsc{i}}}$ with $t_{\smtextsc{s}}$ and $t_{\smtextsc{i}}$ being the time variables associated with the signal and the idler photons respectively, and $Z_{t_{\smtextsc{s}}}\ket{\widetilde{+}}_{\omega_{\smtextsc{s}}}=\ket{\widetilde{-}}_{\omega_{\smtextsc{s}}}$. 
It is convenient to express $\ket{\widetilde{+}}_{\omega_{\smtextsc{x}}}$ and $\ket{\widetilde{-}}_{\omega_{\smtextsc{x}}}$ $(\textsc{x}=\textsc{s,i})$ as
\begin{eqnarray}
\nonumber \ket{\widetilde{+}}_{\omega_{\smtextsc{x}}}=\int \rmd \omega_{\smtextsc{x}}\, &\int \rmd t_{\smtextsc{x}}\, \,\exp\left(-\frac{t_{\smtextsc{x}}^{2}}{2 \kappa_{\smtextsc{x}}^{2}} - \frac{\omega_{\smtextsc{x}}^{2}}{2 (\Delta \omega)^{2}}\right)\\
&\sum_{n}\, e^{i (\omega_{\smtextsc{x}} + n \overline{\omega}) t_{\smtextsc{x}}} \,  \ket{\omega_{\smtextsc{x}} + n \overline{\omega}}, 
\label{eqn:tildePl}
\end{eqnarray}
and
\begin{eqnarray}
\nonumber \ket{\widetilde{-}}_{\omega_{\smtextsc{x}}}=\int \rmd \omega_{\smtextsc{x}}\, &\int \rmd t_{\smtextsc{x}}\, \,\exp\left(-\frac{t_{\smtextsc{x}}^{2}}{2 \kappa_{\smtextsc{x}}^{2}} - \frac{\omega_{\smtextsc{x}}^{2}}{2 (\Delta \omega)^{2}}\right)\\
& \sum_{n} (-1)^{n}\, e^{i (\omega_{\smtextsc{x}} + n \overline{\omega}) t_{\smtextsc{x}}} \,  \ket{\omega_{\smtextsc{x}} + n \overline{\omega}}. 
\label{eqn:tildeMin}
\end{eqnarray}
These expressions follow from the properties of the displacement operator, and the expressions~(B1), (B2), (B7) given in \cite{milman}. Here, $\kappa_{\smtextsc{x}}$ $(\textsc{x}=\textsc{s,i})$ is the standard deviation in $t_{\smtextsc{x}}$.
It follows from~\eref{eqn:tildePl} and \eref{eqn:tildeMin} that
\begin{eqnarray}
\nonumber\ket{\psi^{\prime}}=\int \rmd t \, \int \rmd t^{\prime} \, \int \rmd \omega \, \int \rmd \omega^{\prime} \, \exp \left(- \frac{ t^{2} (\Delta \omega_{p})^{2} + t^{\prime 2} (\Delta \Omega)^{2}}{2}-\frac{\omega^{2} + \omega^{\prime 2}}{2 (\Delta \omega)^{2}}\right) \\
\sum_{n,m} (-1)^{n} e^{i (n \overline{\omega} + \omega) (t+t^{\prime})} e^{i (m \overline{\omega} + \omega^{\prime}) (t-t^{\prime})} \ket{n \overline{\omega} + \omega}\otimes\ket{m \overline{\omega} + \omega^{\prime}},
\label{eqn:psiPrime}
\end{eqnarray}
where $\Delta\omega_{p}$ is the standard deviation in $\omega_{p}$. Integrating over the time variables $t$ and $t^{\prime}$, writing $(\omega_{\smtextsc{s}}=n \overline{\omega} + \omega)$, $(\omega_{\smtextsc{i}}=m \overline{\omega} + \omega^{\prime})$ where $n$, $m$ are integers, and using the fact that $f_{+}$ is a Gaussian function with a standard deviation $\Delta\omega_{p}$ ($\Delta \omega_{p} \ll \Delta \Omega$), it is straightforward to see that~\eref{eqn:psiPrime} can be expressed as 
$\ket{\Psi_{\beta}}$ in ~\eref{eqn:Psi_mipl}, 
 unnormalised. 

\section*{\label{appen:ChronoTomo}Appendix B: Expressions for the chronocyclic tomograms}
\setcounter{section}{2}

We are interested in the time-time slice of the tomograms corresponding to $\ket{\Psi_{\alpha}}$~\eref{eqn:Psi_plpl} and $\ket{\Psi_{\beta}}$~\eref{eqn:Psi_mipl}. As a first step, we calculate the explicit expressions for the states $\ket{\Psi_{\alpha}}$ and $\ket{\Psi_{\beta}}$ in the Fourier transform basis (i.e., time-time basis) using $f_{+}(\omega_{\smtextsc{s}}+\omega_{\smtextsc{i}})= \delta(\omega_{\smtextsc{s}}+\omega_{\smtextsc{i}}-\omega_{p})$ in~\eref{eqn:Psi_plpl} and \eref{eqn:Psi_mipl}. The biphoton state $\ket{\Psi_{\alpha}}$ in the time-time basis is given by
\begin{eqnarray}
\nonumber \ket{\Psi_{\alpha}}= \frac{1}{\sqrt{\mathcal{M}_{\alpha}\tau_{\smtextsc{p}}}} \,\int \rmd t_{\smtextsc{s}} \,\int \rmd t_{\smtextsc{i}} \, \exp \left( -\frac{(t_{\smtextsc{i}}-t_{\smtextsc{s}})^{2} \, (\vardel\omega)^{2} \, (\vardel\Omega)^{2}}{4\: ((\vardel\omega)^{2} + (\vardel\Omega)^{2})} \right) \\
\times \left[\mathcal{F}(t_{\smtextsc{i}} - t_{\smtextsc{s}})\right]^{2} \ket{t_{\smtextsc{s}};t_{\smtextsc{i}}},
\label{eqn:Psi_plpl_tt}
\end{eqnarray}
and the time-time slice $w^{\alpha}(t_{\smtextsc{s}};t_{\smtextsc{i}})$ corresponding to $\ket{\Psi_{\alpha}}$ is
\begin{eqnarray*}
w^{\alpha}(t_{\smtextsc{s}};t_{\smtextsc{i}}) = \frac{1}{\mathcal{M}_{\alpha}\tau_{\smtextsc{p}}} \, \exp \left( -\frac{(t_{\smtextsc{i}}-t_{\smtextsc{s}})^{2} \, (\vardel\omega)^{2} \, (\vardel\Omega)^{2}}{2\: ((\vardel\omega)^{2} + (\vardel\Omega)^{2})} \right) 
 \Big\vert\mathcal{F}(t_{\smtextsc{i}} - t_{\smtextsc{s}}) \Big\vert^{4},
\end{eqnarray*}
where $\tau_{\smtextsc{p}}=1$s (introduced for dimensional purposes), 
\begin{eqnarray*}
\mathcal{F}(t_{\smtextsc{i}} - t_{\smtextsc{s}}) = \sum\limits_{n} \exp \left( \frac{i (t_{\smtextsc{i}} - t_{\smtextsc{s}})\, n \, \overline{\omega} (\vardel\Omega)^{2}}{2 ((\vardel\omega)^{2} + (\vardel\Omega)^{2})} \right),
\end{eqnarray*}
and the normalisation constant is
\begin{eqnarray}
\mathcal{M}_{\alpha}= \frac{\pi}{\mu_{0}} \sum_{m,n,m',n'} \exp \left( -\frac{(n-n'+m'-m)^{2}\, \overline{\omega}^{2}\, (\vardel\Omega)^{2}}{2\, (\vardel\omega)^{2}\: ((\vardel\omega)^{2} + (\vardel\Omega)^{2})} \right),
\label{eqn:NormConst2Alph}
\end{eqnarray}
where $\mu_{0}=\left(\frac{\pi (\vardel \Omega)^{2} (\vardel \omega)^{2}}{2 ((\vardel \Omega)^{2} + (\vardel \omega)^{2})}\right)^{1/2}$.

Similarly, the other biphoton state $\ket{\Psi_{\beta}}$ in the time-time basis is given by
\begin{eqnarray}
\nonumber \ket{\Psi_{\beta}}= \frac{1}{\sqrt{\mathcal{M}_{\beta}\tau_{\smtextsc{p}}}} \,\int \rmd t_{\smtextsc{s}} \,\int \rmd t_{\smtextsc{i}} \, \exp \left( -\frac{(t_{\smtextsc{i}}-t_{\smtextsc{s}})^{2} \, (\vardel\omega)^{2} \, (\vardel\Omega)^{2}}{4\: ((\vardel\omega)^{2} + (\vardel\Omega)^{2})} \right)\\
\times \mathcal{G}(t_{\smtextsc{i}} - t_{\smtextsc{s}}) \,
\mathcal{F}(t_{\smtextsc{i}} - t_{\smtextsc{s}}) \ket{t_{\smtextsc{s}};t_{\smtextsc{i}}},
\label{eqn:Psi_mipl_tt}
\end{eqnarray}
and the time-time slice corresponding to $\ket{\Psi_{\beta}}$ is
\begin{eqnarray*}
w^{\beta}(t_{\smtextsc{s}};t_{\smtextsc{i}}) = \frac{1}{\mathcal{M}_{\beta}\tau_{\smtextsc{p}}} \, \exp \left( -\frac{(t_{\smtextsc{i}}-t_{\smtextsc{s}})^{2} \, (\vardel\omega)^{2} \, (\vardel\Omega)^{2}}{2\: ((\vardel\omega)^{2} + (\vardel\Omega)^{2})} \right)  \Big\vert \mathcal{G}(t_{\smtextsc{i}} - t_{\smtextsc{s}}) \, \mathcal{F}(t_{\smtextsc{i}} - t_{\smtextsc{s}}) \Big\vert^{2},
\end{eqnarray*}
where,
\begin{eqnarray*}
\mathcal{G}(t_{\smtextsc{i}} - t_{\smtextsc{s}}) = \sum\limits_{n} (-1)^{n} \exp \left( \frac{i (t_{\smtextsc{i}} - t_{\smtextsc{s}})\, n \, \overline{\omega} (\vardel\Omega)^{2}}{2 ((\vardel\omega)^{2} + (\vardel\Omega)^{2})} \right),
\end{eqnarray*} 
and the normalisation constant $\mathcal{M}_{\beta}$ is essentially the same as $\mathcal{M}_{\alpha}$ with an extra factor of $(-1)^{n+n'}$ within the summation in~\eref{eqn:NormConst2Alph}.

\section*{References}
\bibliography{references}

\end{document}